\documentclass[12pt,preprint]{aastex}
\input{epsf}
\begin{document}

\def\msun{\hbox{${\cal{M}}_{\odot}$}}
\def\massA{\hbox{${\cal{M}}_A$}}
\def\massB{\hbox{${\cal{M}}_B$}}
\def\massAB{\hbox{${\cal{M}}_{A+B}$}}

\slugcomment{accepted for {\it The Astronomical Journal}, 11/04/08}

\shorttitle{Multiplicity of Massive Stars}
\shortauthors{Mason et al.}

\title{The High Angular Resolution Multiplicity of Massive Stars}

\author{Brian D. Mason\altaffilmark{1}, William I. Hartkopf\altaffilmark{1}}
\affil{U.\ S.\ Naval Observatory, \\
3450 Massachusetts Avenue, NW, Washington, DC, 20392-5420; \\ 
Electronic mail: (bdm, wih)@usno.navy.mil}

\author{Douglas R. Gies, Todd J. Henry}
\affil{Center for High Angular Resolution Astronomy, \\ 
Department of Physics and Astronomy, \\
Georgia State University, P. O.\ Box 4106, Atlanta, GA, 30302-4106; \\ 
Electronic mail: (gies, thenry)@chara.gsu.edu}

\author{John W. Helsel}
\affil{Department of Physics, Furman University, \\
3300 Poinsett Hwy., Greenville, SC 29613; \\ 
Electronic mail: john.helsel@furman.edu}

\altaffiltext{1}{Visiting Astronomer, Kitt Peak National Observatory and 
Cerro Tololo Inter-American Observatory, National Optical Astronomy 
Observatory, operated by Association of Universities for Research in 
Astronomy, Inc.\ under contract to the National Science Foundation.}


\begin{abstract}

We present the results of a speckle interferometric survey of Galactic 
massive stars that complements and expands upon a similar survey made over a
decade ago. The speckle observations were made with the KPNO and CTIO 4~m 
telescopes and USNO speckle camera, and they are sensitive to the detection
of binaries in the angular separation regime between $0\farcs03$ and 
$5\arcsec$ with relatively bright companions ($\Delta V < 3$). We report on
the discovery of companions to 14 OB stars. In total we resolved companions
of 41 of 385 O-stars (11\%), 4 of 37 Wolf-Rayet stars (11\%), and 89 of 139
B-stars (64\%; an enriched visual binary sample that we selected for future 
orbital determinations). We made a statistical analysis of the binary 
frequency among the subsample that are listed in the Galactic O Star Catalog
by compiling published data on other visual companions detected through 
adaptive optics studies and/or noted in the Washington Double Star Catalog 
and by collecting published information on radial velocities and 
spectroscopic binaries. We find that the binary frequency is much higher 
among O-stars in clusters and associations compared to the numbers for field
and runaway O-stars, consistent with predictions for the ejection processes 
for runaway stars. We present a first orbit for the O-star $\delta$ Orionis,
a linear solution of the close, apparently optical, companion of the O-star 
$\iota$ Orionis, and an improved orbit of the Be star $\delta$ Scorpii. 
Finally, we list astrometric data for another 249 resolved and 221 
unresolved targets that are lower mass stars that we observed for various 
other science programs. 
\end{abstract}



\setcounter{footnote}{1} 

\section{Introduction}

Massive stars appear to love company. There is growing evidence that the 
incidence of binary and multiple stars among the massive O- and B-type stars
is much larger than that for solar type stars (see Zinnecker \& Yorke 2007
and references therein). This difference in multiplicity properties may 
ultimately reflect differences in the star formation process between massive
and low mass stars. For example, while low mass stars may lose angular 
momentum by magnetic and disk-related processes, it may be that these are 
ineffective in massive star formation because of the very short timescale of
formation. Instead, the initial angular momentum of the natal cloud may end 
up (through a variety of processes) in the orbital angular momentum of 
binaries among the more massive stars (Bate et al.\ 2002; Zinnecker \& Yorke
2007; Gies 2007). 

The observational evidence for the high incidence of binaries among the 
massive stars comes from spectroscopic investigations of short-period 
systems and high angular resolution measurements of longer-period (and wide)
binaries. We made one of the most comprehensive surveys of the bright, 
Galactic O-type stars in a speckle interferometric study made in 1994 with 
the NOAO 4~m telescopes in both the northern and southern hemispheres (Mason
et al.\ 1998). This investigation considered both speckle measurements and 
published data on radial velocity measurements to determine the overall 
binary properties among stars in clusters and associations, field O-stars, 
and runaway O-stars. The results indicated a much higher incidence of 
binaries among O-stars in clusters and associations, and we suggested that 
the true binary frequency may reach 100\% among cluster stars once account 
is made for the observational bias against detection of binaries with 
periods larger than those found spectroscopically but smaller than those 
found through high angular resolution measurements. This work was 
complemented by similar speckle interferometric surveys of Wolf-Rayet stars
(Hartkopf et al.\ 1999) and Be stars (Mason et al.\ 1997a). 

Ten years later (and armed with an improved detector) we decided it was an 
opportune time for follow up and expanded speckle observations. A second 
epoch survey is desirable for a number of reasons. Some systems observed in
1994 may have been situated in orbital phases of close separation, and hence
were unresolved. Since the systems detectable by speckle correspond to
periods of decades for massive stars, it is important to repeat the survey
after a similar time span. Furthermore there are a significant number of 
specific systems where new observations are particularly important. For 
example, there are several cases where a triple is indicated by 
spectroscopy, but we have yet to resolve the wide system (e.g., $\delta$ 
Cir; Penny et al.\ 2001). The placement of many of the very hot, O2 and O3 
stars in the Hertzsprung-Russell Diagram suggests that they are very 
massive because they are so bright, but sometimes this extreme luminosity 
is instead due to the presence of a companion (Nelan et al.\ 2004; 
Niemela \& Gamen 2005; Ma\'{i}z-Apell\'{a}niz et al.\ 2007). The massive 
binaries in the Orion Trapezium detected in the near-IR by Schertl et al.\
(2003) have separations that are within the resolution limit of a 4~m 
telescope, and detection or not of these companions at another wavelength 
can help set limits on the magnitude difference $\Delta m$, the color, and 
hence object type. For systems with two speckle measurements, a third one 
may allow the motion to be recognized as either linear or non-linear (i.e.,
Keplerian), indicating whether the pair is optical or physical. This is 
extremely important in the case of $\iota$~Ori, where dynamical analysis 
(Gualandris et al.\ 2004) of this complex runaway system virtually requires
that the speckle companion (first reported in Mason et al.\ 1998 at only 
$0\farcs11$ separation) be optical rather than physical. Finally, such high 
angular resolution measurements can provide direct astrometric orbits (for 
the nearby systems) and hence mass measurements for binaries that are 
clearly non-interacting (Vanbeveren et al.\ 1998). These provide fundamental
data on the masses and other properties of the most massive stars. 

For all these reasons, we embarked on a new survey of speckle interferometry
measurements of the massive stars that were mainly selected from the 
Galactic O-star Catalog (Ma\'{i}z-Apell\'{a}niz \& Walborn 2004). We 
describe the observational program in \S2 and outline the main tabular 
results in \S3. We use these results to reassess the binary properties of 
the O-stars in \S4, and then we discuss the the results for specific targets
in \S5. The observational program included a significant number of other, 
less massive stars, and these measurements and several updated astrometric 
orbits are given in Appendices A and B, respectively. 

\section{Instrumentation and Calibration}

The instrument used for most of these observations was the USNO speckle
interferometer, described most recently by Hartkopf et al.\ (2008). Three
different filters were selected, all having approximately the same central
wavelength but with different full width at half maximum (FWHM) band passes. 
Of these, two are standard filters (Str\"{o}mgren $y$, 550$\pm$24 nm, and 
Johnson $V$, 545$\pm$85 nm). An intermediate filter, designated USNO green 
(560$\pm$45 nm), was also used. While the Johnson $V$ allows the camera to
observe much fainter targets, the resolution limit is degraded to about 
0\farcs05. Both of the other filters reached the goal resolution limit of 
0\farcs03. We selected a filter for each target with a band width suitable 
to the magnitude of the star and which allowed us to detect an adequate 
number of speckles. These resolution limit values are most significant when
no companion was detected (Tables 3 and B2). Instances when the wider 
Johnson filter was used are indicated with a note to these tables.

Observations of northern hemisphere objects were obtained with the KPNO 4~m 
Mayall Reflector during the period 2005 November 8 -- 13; southern 
hemisphere pairs were observed at the CTIO 4~m Blanco Reflector during the 
period 2006 March 9 -- 13. Atmospheric conditions during both runs were 
exceptional, with excellent transparency and significant periods of 
sub-arcsecond seeing with both telescopes, especially at Cerro Tololo. On 
these two runs, 1876 observations were obtained, resulting in 652 measures 
of double stars and 1050 high-quality observations where a pair was 
definitively not seen. The remaining observations were of insufficient 
quality for a definitive measure. Additional observations of massive stars 
were obtained during other 4~m observing runs as listed below.

Calibration of the KPNO data was determined through the use of a double-slit
mask placed over the ``stove pipe'' of the 4~m telescope during observations
of a bright known-single star (as described in Hartkopf et al.\ 2000). This 
application of the well known experiment of Young allowed determination of 
scale and position angle zero point without relying on binaries themselves 
to determine calibration parameters. Multiple observations through the slit
mask (during five separate KPNO runs from 2001 to 2008) yielded mean errors
of 0\fdg11 in the position angle zero point and 0.165\% in the scale error. 
These ``internal errors'' are undoubtedly underestimates of the true errors
of these observations.  Plate scales for the five Kitt Peak runs, 2001 
January, 2001 July, 2005 November, 2007 August, and 2008 June, were found to
be 0.01257, 0.01282, 0.01095, 0.01090, and 0.01096 arcseconds per pixel, 
respectively. While the camera remained the same for all five runs, the 
latter three were obtained with a newer computer and frame grabber and a 
different set of microscope objectives. The effective field-of-view for 
detection of binaries is 1\farcs5 for nominal conditions and 3\farcs0 when
the targets are fainter and a lower microscope objective is used with the
Johnson $V$ filter. Wider, easily detected pairs can be accomodated with a 
larger 6\farcs0 field-of-view with a low power microscope objective and 
2$\times$2 pixel averaging.

Since the slit-mask option was not available on the CTIO 4~m telescope, we 
calibrated the southern hemisphere data using observations of numerous 
well-observed, wide, and equatorially located binaries that we observed with
both the KPNO and CTIO telescopes. Published orbital elements for these 
pairs were updated as needed using the recent KPNO measures, then predicted 
$\rho$ and $\theta$ values from those orbits deemed of sufficiently high 
quality were used to determine the CTIO scale and position angle zero point.
The calibration errors for these southern observations were (not 
surprisingly) considerably higher than those achieved using the slit mask. 
Mean errors for three CTIO runs from 2001 to 2006 were 0\fdg67 in position 
angle and 1.44\% in scale. Plate scales for the three Cerro Tololo runs, 
2001 January, 2001 July, and 2006 March, were 0.01262, 0.01253, and 0.01084
arcseconds per pixel, respectively. The differences are attributable to 
changes in equipment as described above. The field-of-view was comparable 
for the southern and northern observations. 

Speckle Interferometry is a technique which is very sensitive to changes in
observing conditions, particularly coherence length ($\rho_0$) and time 
($\tau_0$). These are typically manifested as a degradation of detection 
capability close to the resolution limit or at larger magnitude differences.
To ensure we are reaching our desired detection thresholds, a variety of 
systems with well-determined morphologies and magnitude differences were 
observed throughout every observing night. In all cases, the observations of
these test objects indicated that our measurements met or exceeded these 
thresholds as indicated in Figure~1.

\begin{figure}[t]
\epsfxsize 3.0in
\centerline{\epsffile{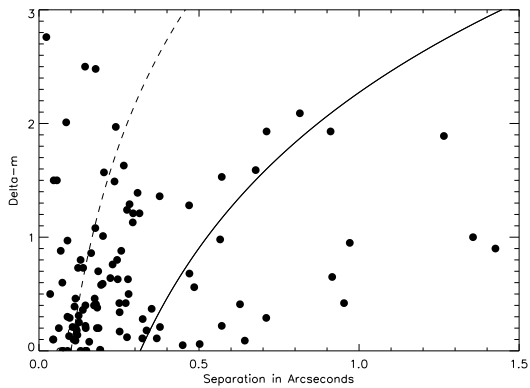}}
\caption{Plot of separation ($\rho$) vs. magnitude difference ($\Delta m$) 
for pairs observed.  The separations are direct measurements from Tables 2 
and A1 while $\Delta m$ is the tabulated value from Washington Double Star 
Catalog (WDS; Mason et al.\ 2001). The curved lines indicate the measure of
difficulty relationship of \"{O}pik (1924) as modified by Heintz (1978a). 
The $\rho-\Delta$m combinations below the solid line are considered 
completely known. Those above the dashed line are considered virtually 
unknown. Filled circles are those objects observed to investigate detection
capabilities. The quality of data exceeded expectation. The most challenging
object, 22430$+$3013 or BLA~~11Aa, at upper left, has a measured separation 
of 0\farcs051 and a magnitude difference of 2.76 (as determined by the Mark 
III optical interferometer; Hummel et al.\ 1998).}
\label{fig1}
\end{figure}

\placefigure{fig1}
 
\section{Results}

The target list consists of the original sample of O-stars from Mason et 
al.\ (1998), additional O-stars from the catalog of Ma\'{i}z-Apell\'{a}niz 
\& Walborn (2004), WR stars, and B-stars. The B-star sample includes 
candidates for orbit and mass determination, Pleiades cluster members 
observed previously (Mason et al.\ 1993b), and Be stars (Mason et al.\ 
1997a). A number of low mass targets were also observed that are discussed 
in Appendix A. 

Table~1 presents coordinates and magnitude information from 
CDS\footnote{Magnitude information is from the Aladin Sky Atlas, operated at
CDS, Strasbourg, France.} for all those binaries which are resolved or 
measured for the first time. Column 1 gives the coordinates of the primary 
of the pair. Column 2 lists the discoverer designation number (with WSI = 
Washington Speckle Interferometry), and column 3 gives an alternative 
designation. Column 4 provides the spectral classification, and column 5 the
combined visual magnitude. Finally, column 6 refers to notes below the 
table. 

\placetable{tab1}

Table~2 lists the astrometric measures of the observed massive binaries. 
They are subdivided into four groups consisting of the original 1998 
sample of O-stars, the newer set of O-stars, WR stars, and B-stars. The 
first three columns identify the system by providing the epoch-2000 
coordinates, discovery designation, and an alternate designation. Columns 4
through 6 give the epoch of observation (expressed as a fractional Besselian
year), the position angle $\theta$ (in degrees), and the separation $\rho$ 
(in seconds of arc). Note that the position angle has not been corrected for
precession, and is thus based on the equinox for the epoch of observation. 
Objects whose measures are of lower quality are indicated by colons 
following the position angle and separation. These lower-quality 
measurements may be due to one or more of the following factors: close 
separation, large $\Delta m$, one or both components very faint, a large 
zenith distance at the time of observation, and poor seeing or transparency.
They are included primarily because they confirm an earlier observation or 
because a long time has elapsed since the last measurement. Column 7 
provides the $V$-band magnitude difference. This is usually a catalog value
from the WDS (Mason et al.\ 2001), although for new pairs and some other 
infrequently measured interferometric pairs it is a crude value based upon 
the strengths of the secondary peak and ``anti-peak'' in Fourier Transform 
space, as seen in the generated directed vector autocorrelations (Bagnuolo
et al.\ 1992). Differential magnitudes were ``calibrated'' by direct 
comparison with other pairs of known magnitude difference and are probably 
accurate to $\pm$0.5 mag. Column 8 indicates the number of observations used
to derive the mean position (usually 1). For systems with orbits, the 
observed minus calculated residuals $O-C$ for both $\theta$ and $\rho$ are 
given in columns 9 and 10 according to the orbit whose reference is given in
column 11. Finally, column 12 refers to specific notes for these systems. 
Some measures from other KPNO/CTIO 4~m runs are noted and listed here and in
Table~3. 

\placetable{tab2}

Table~3 provides a complete list of single star observations for the massive
star sample. The precise coordinate ($\alpha$,$\delta$) is given in column 
1, while columns 2 -- 4 list various designations. A code for the massive 
star sub-sample is given in column 5, and the Besselian date of observation 
appears in column 6. Column 7 indicates with a K or C if the 4~m telescope 
used for the observation is the Mayall reflector at KPNO (K) or the Blanco 
reflector at CTIO (C). Finally, column 8 provides notes for the stars.

\placetable{tab3}


\section{Binary Frequency of O-type Stars}

It is important to consider the environment of massive stars in the 
determination of binary frequency. While most massive stars are found close
to their birthplaces in stellar clusters and OB associations, there are 
significant numbers of ``field'' O-stars (which have no apparent nearby 
cluster; de Wit et al.\ 2005) and ``runaway'' O-stars (high velocity or 
remote from the Galactic plane; Gies \& Bolton 1986) that were probably 
ejected from clusters. The ejection process may have involved close 
gravitational encounters of binaries and/or supernovae explosions in 
binaries (Hoogerwerf et al.\ 2000; Zinnecker \& Yorke 2007), and such 
ejected stars will generally be single objects. In our original speckle 
survey (Mason et al.\ 1998), we found that indeed the binary fraction 
decreased among field and runaway O-stars compared to those in clusters and
associations. 
 
Here we revisit the question of the binary frequency of massive stars based 
upon the results from our speckle interferometric survey. We will restrict 
our sample to the O-stars appearing in the Galactic O Star Catalog of 
Ma\'{i}z-Apell\'{a}niz \& Walborn (2004), since we now have speckle data for
360 of the 370 stars in the catalog. These stars and their binary properties
are listed in Table~4, using the same names and order (based upon increasing
Galactic longitude) as given in the Galactic O Star Catalog.

\placetable{tab4}

The second column of Table~4 gives a code for the short period, 
spectroscopic binary status based upon a literature search through 2008 
August. These codes are similar to those adopted by Mason et al.\ (1998),
and we use a ``SB'' prefix for known or probable spectroscopic binaries, a 
``C'' for constant velocity stars, and a ``U'' for stars of unknown status 
(usually with fewer than four radial velocity measurements). The SB stars 
with a published orbit have an ``O'' suffix attached to the code and a 
middle numeral that represents the number of spectral components identified.
Usually a code of ``SB2O'' represents a double-lined spectroscopic binary, 
but we also apply it to cases such as QZ~Car = HD~93206 that consists of two
single-lined binaries in a quadruple system. The ``SB3O'' code is applied to
triple systems where a third, stationary, spectral component is visible at 
the greatest velocity separation of the double-lined system.  An ``E'' 
suffix denotes the presence of orbital flux variations (eclipses or 
ellipsoidal variations), and the ``SBE'' code indicates that we know the 
star is a binary from the light curve but no spectroscopic investigation 
exists yet. The suspected spectroscopic binaries are coded by ``SB2?'' 
(where observers report line doubling) and ``SB1?'' (where the range in 
measured radial velocity exceeds 35 km~s$^{-1}$). The most recent published
reference is indicated by the SAO/NASA Astrophysics Data System 
bibliographic code in column 6 of Table~4. 

The number of angularly resolved components is given in column 3 of Table~4. 
This represents the sum of the number of close components found by speckle 
interferometry, wider and fainter components found by Turner et al.\ (2008b)
in an adaptive optics survey, and other (usually wider) components listed in
the WDS (Mason et al.\ 2001). These sources were supplemented by detailed 
studies of specific stars or clusters, such as $\zeta$ Ori (long baseline 
optical interferometry; Hummel et al.\ 2000), Trumpler 14 ({\it HST} FGS; 
Nelan et al.\ 2004), the Orion Trapezium (infrared single aperture 
interferometry; Petr et al.\ 1998; Simon et al.\ 1999; Weigelt et al.\ 1999;
Kraus et al.\ 2007), and NGC~6611 (Duch\^{e}ne et al.\ 2001). A quotation 
mark in this column indicates that the star is a member of a visual system 
whose primary component also appears in the table (usually just above or 
below such an entry), and a colon marks those stars that lack speckle 
observations. Note that a large number of visual components may indicate 
that the star resides at the center of a dense cluster.  

Column 4 of Table~4 associates the star with the field or the name of the 
home cluster, while column 5 lists whether or not the star is considered to 
be a runaway object. These determinations come directly from the Galactic O 
Star Catalog (Ma\'{i}z-Apell\'{a}niz \& Walborn 2004) with new runaway 
identifications noted by Mdzinarishvili (2004) and de Wit et al.\ (2005). 
Note that some runaway stars can be traced to a cluster of origin, so that 
they will be assigned to that cluster in column 4. 

The binary statistics derived from Table~4 are summarized in Table~5 (an 
updated version of Table~3 from Mason et al.\ 1998). We caution that the 
sample is magnitude limited (and therefore biased to more luminous stars) 
and incompletely surveyed (for example, the Turner et al.\ (2008b) adaptive 
optics work is limited to stars with declination $> -42^\circ$). The stars 
are grouped into cluster/association, field, and runaway categories to 
compare the binary properties. For the immediate purpose of this work, we 
simply assigned any star that was not a field or runaway object to the 
cluster/association category. This includes stars described as more distant
than some foreground cluster, since such stars generally reside along a 
spiral arm of the Galaxy where cluster membership is common. The top 
section of Table~5 summarizes the visual multiplicity properties of each 
category for the 347 unique, visual systems in the Galactic O Star Catalog.
The results are presented in rows that correspond to the sum based upon the
number of visual components $n$ found. We divide the sample into single and 
multiple groups in determining the percentages without and with companions 
(making the tacit assumption that most of the visual companions are 
gravitationally bound and not line of sight optical companions).  

\placetable{tab5}

The middle section of Table~5 presents the corresponding sums for the 
spectroscopic binary properties for all 370 entries in the Galactic O Star 
Catalog. The percentages for each subgroup represent fractions with the 
unknown ``U'' status objects excluded from the totals. Finally, the lower 
section in Table~5 shows the percentages for the presence of any companion 
(spectroscopic or visual) again excluding the stars with unknown 
spectroscopic status. 

The results from this larger sample tend to confirm the trend found by Mason
et al.\ (1998) that the binary frequency is lower among field and runaway 
stars than that found in the cluster/association group. The binaries found 
among the runaway stars tend to be close systems with nearly equal mass 
components (HD~1337, $\iota$~Ori, Y~Cyg) and binaries with neutron star 
companions (HD~14633, HD~15137, X~Per, HD~153919). The former group are 
predicted to be infrequently ejected in close gravitational encounters 
(Leonard \& Duncan 1990) while the latter are the result of a supernova 
explosion in a binary, so both processes must contribute to the ejection of
massive stars from clusters. A number of runaways have visual companions 
that must be optical, chance alignments, since the ejection processes are 
too energetic for soft, wide binaries to survive.   

The binary statistics for the cluster and association group offer us the 
best estimate of the binary properties at birth (before dynamical and 
stellar evolution processes alter the statistics). Our results indicate that
most O-stars (and by extension most massive stars) are born in binary or 
multiple star systems. This result is especially striking since those 
binaries with orbital periods too long for easy spectroscopic detection and
too short for direct angular resolution are absent from the totals, so the 
fractions reported here are clearly lower limits for the binary frequency.
Thus, the processes that lead to the formation of massive stars strongly 
favor the production of binary and multiple star systems. 

\section{Individual Systems}

\subsection{$\iota$ Ori = CHR~250}

The complex dynamical relationship of AE Aur, $\mu$ Col, and $\iota$ Ori 
is one of the best examples of a binary-binary collision (Gies \& Bolton 
1986; Leonard \& Duncan 1990; Leonard 1995; Clarke \& Pringle 1992). As
$\iota$ Ori is a known close pair ($P = 29.13376$ d; Marchenko et al.\ 
2000), the much wider speckle component would be hierarchical if physical, 
with an estimated period of at least 40~y (Gualandris et al.\ 2004). As the
high energy needed to eject AE Aur and $\mu$ Col with their runaway 
velocities seemed inconsistent with the less energetic dynamical interaction
required for the CHR~250 pair to remain bound, Gualandris et al.\ (2004) 
postulated that this pair was non-physical, despite their close proximity. 
Figure~2 shows a least-squares, linear fit\footnote{See Hartkopf, W. I., 
Mason, B. D., Wycoff, G. L, \& Kang, D. ``Catalog of Rectilinear Elements''
{\tt http://ad.usno.navy.mil/wds/lin1.html}.} to the published data (Mason 
et al.\ 1998 and Table~2). The data are also consistent with a long-period 
orbit, but much longer than $\approx$ 40~y.

\begin{figure}[!ht]
\epsfxsize 5.0in
\centerline{\epsffile{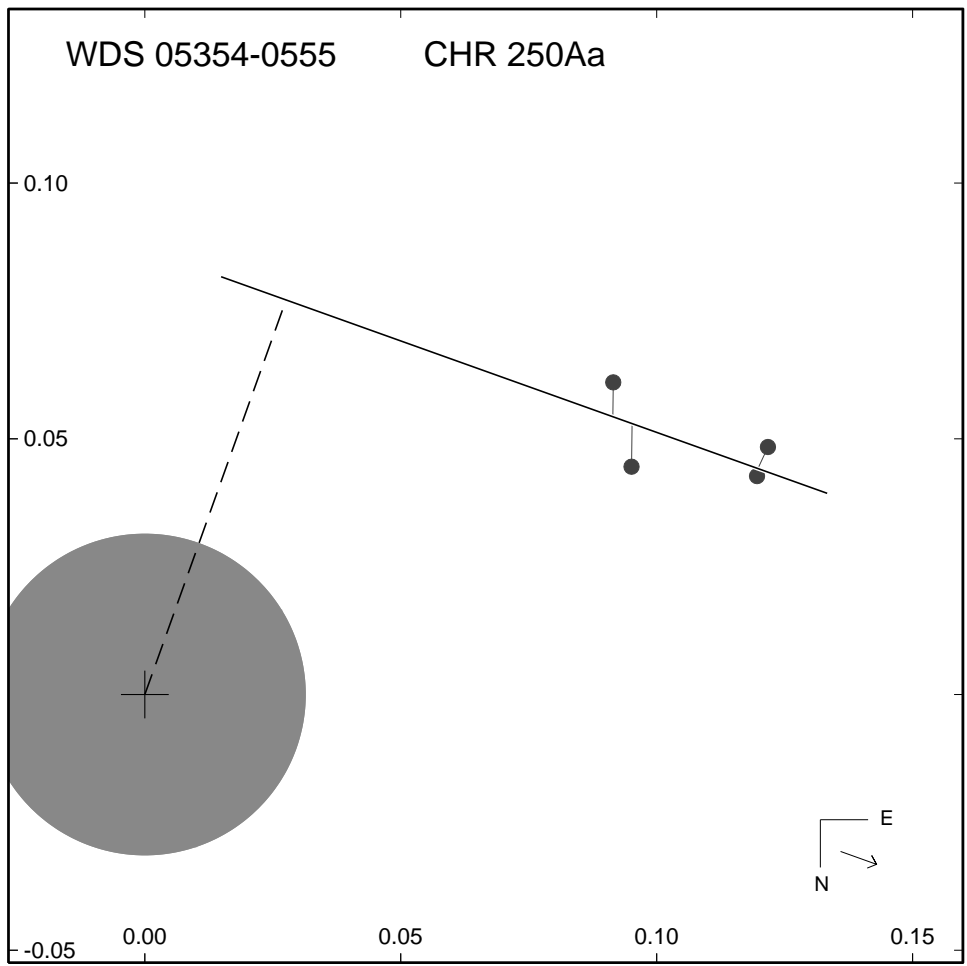}}
\caption{The relative motion of the components of CHR 250 = $\iota$ Ori. The
straight line is a rectilinear fit to the four measures (two from Mason et 
al.\ 1998, two from Table~2), indicating motion to the ENE. The shaded 
circle indicates the $\sim$30 mas resolution limit of a 4~m telescope, while
the dashed line indicates the closest separation of the two stars assuming 
their relative motion is rectilinear. The stars appear to have reached a 
closest separation of 82$\pm$5 mas in 1969.7. Of course, the entire time 
span of observations of this pair is only about 11.5 years; we may instead 
be observing only a small arc of a long-period orbit.}
\label{fig2}
\end{figure}

\placefigure{fig2}
 
\subsection{$\delta$ Ori = HEI~42}

We present a first orbit for the wide component of this triple system that 
is based on all available published data and the new measures listed in 
Table~2. The previous measurements were extracted from the WDS (Mason et 
al.\ 2001) and were weighted following the precepts of Hartkopf et al.\ 
(2001b). The orbital elements were determined with an iterative 
three-dimensional grid-search algorithm (Seymour et al.\ 2002). The seven 
orbital elements are presented in Table~6: $P$ (period, in years), $a$ 
(semimajor axis, in arcseconds), $i$ (inclination, in degrees), $\Omega$ 
(longitude of the node, equinox 2000, in degrees), $T$ (epoch of periastron
passage, in fractional Besselian year), $e$ (eccentricity), and $\omega$ 
(longitude of periastron, in degrees). A future ephemeris for the period 
2010 to 2018, in 2-year increments, is provided in Table~B2. The orbit is 
illustrated in Figure~3.

\placetable{tab6}

Due to the preliminary nature and incomplete phase coverage of the orbital 
fit, the errors are large and difficult to quantify. It is entirely possible
that the companion may continue moving to the southeast for longer than the 
orbit plot and ephemeris would indicate. The orbit here then may prove 
wildly erroneous, however, it does serve to highlight the need for periodic 
monitoring of the pair to verify the orbit predictions. The preliminary 
orbit indicates a total mass of $32 M_\odot$ for a distance of 414 pc 
(Menten et al.\ 2007).

The A component is itself a close binary with an orbital period of about 5.7
days (see Harvin et al.\ 2002 for a thorough analysis of the close pair). 
Curiously, the preliminary orbital period, 201 y, is close to the derived 
apsidal period of the close binary (227$\pm$37 y, Monet 1980; 225$\pm$27 y, 
Harvey et al.\ 1987). 

\begin{figure}[!ht]
\epsfxsize 5.0in
\centerline{\epsffile{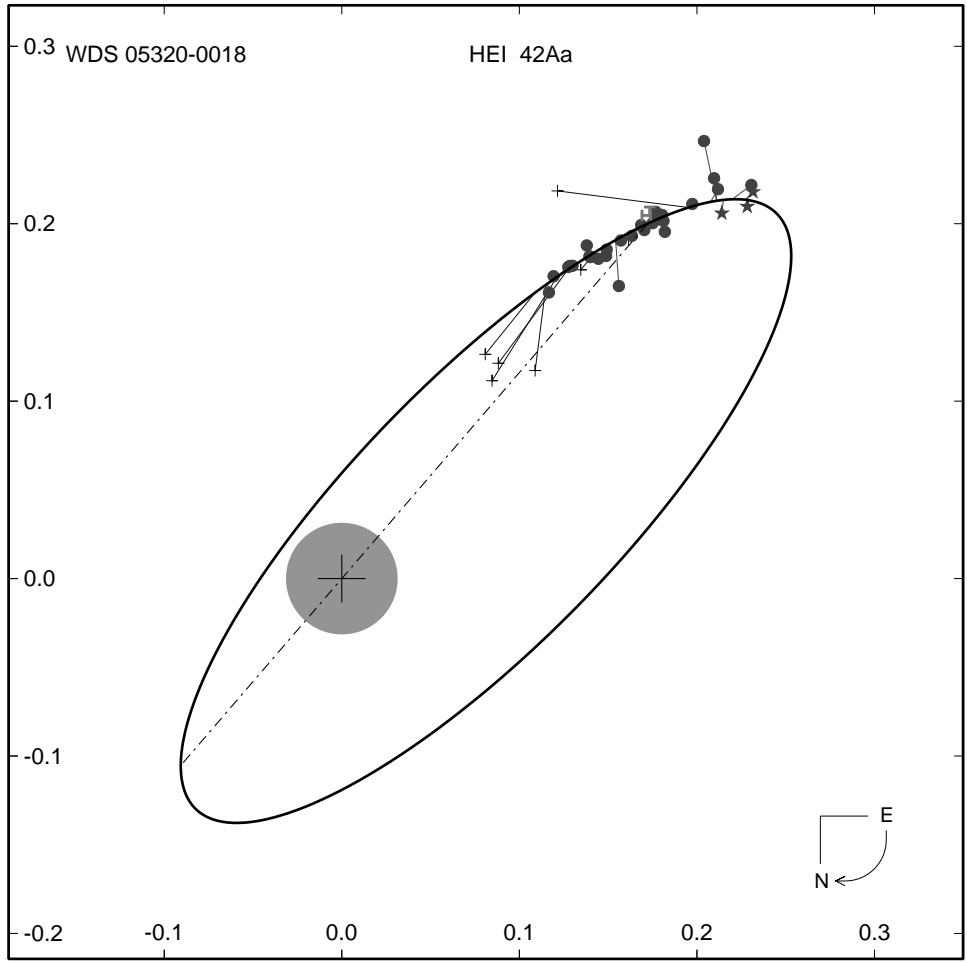}}
\caption[]{Preliminary orbit for $\delta$ Ori. The figure shows the relative
motion of the secondary about the primary (indicated by a large ``plus'' 
sign); the $x$ and $y$ scales are in arcseconds. The solid curve represents 
the new orbit determination. The dot-dash line indicates the line of nodes. 
The three measures from Table~2 are shown as filled stars and all other high
resolution measurements as filled circles. Micrometer measures are indicated
by small plus signs. All measurements are connected to their predicted 
positions on the orbit by ``$O-C$'' lines. The direction of motion is 
indicated on the north-east orientation in the lower right of the plot. The 
gray filled circle centered on the primary represents that region where the 
pair is too close to be resolved by speckle interferometry with a 4~m 
telescope.}
\label{fig3}
\end{figure}

\placefigure{fig3}
 
\subsection{$\delta$ Sco = LAB~3}

Bedding (1993) published the first set of orbital elements for $\delta$ Sco,
followed a few years later by an updated solution from Hartkopf et al.\ 
(1996). Both solutions were based solely on interferometric data (speckle 
interferometry plus two measures made using aperture masking). 
Miroshnichenko et al.\ (2001) obtained complementary radial velocity data 
which tied down $T$ quite precisely and also gave a more accurate estimate 
of the eccentricity, while adopting the values for period and semimajor axis
obtained by Hartkopf et al.\ (1996).

Since the 1996 solution, observations have covered over one additional 
revolution. Published data include a speckle measure by Horch et al.\ (1999)
and one measure by Hipparcos (ESA 1997). This paper includes new speckle 
measures from the Kitt Peak and Cerro Tololo 4~m telescopes, the Mount 
Wilson 100~inch, and the USNO (Flagstaff Station) 61~inch, as well as 
unpublished KPNO and CTIO 4~m observations made with the CHARA speckle 
camera. A new orbital solution was determined, utilizing all available 
interferometric data and adopting the $T$ and eccentricity values of 
Miroshnichenko et al.\ (2001). Elements from this new orbit, as well as the
previously published solutions are given in Table~7; future ephemerides for
the new orbit are given in Table~B2. The new solution and all data used in 
its determination are shown in Figure~4. Here speckle data from this paper 
(Table~2) are shown as filled stars, while other interferometry measures are
indicated by filled circles; the Hipparcos measure is shown as a letter 
``H''. Measures are connected to their predicted locations along the orbit 
by ``$O-C$'' lines; dotted lines indicate measures given zero weight in the 
final orbital solution. The dot-dash line indicates the line of nodes and 
the shaded circle surrounding the origin indicates the Rayleigh separation 
limit for a 4~m telescope. At two epochs in early 1990, observations 
obtained with the KPNO and CTIO 4~m telescopes did not resolve the pair; 
these are indicated by dotted O-C lines from the origin to their predicted 
locations along the orbit. According to the orbital solution, these 
observations should have been marginally resolved. However, given a 
magnitude difference $\Delta m >2$ mag, the lack of resolution so close to 
the Rayleigh limit is not at all surprising. The total mass of the system is
approximately $27 M_\odot$ for a distance of 140~pc (Shatsky \& Tokovinin 
2002). 

\placetable{tab7}

\begin{figure}[!ht]
\epsfxsize 5.0in
\centerline{\epsffile{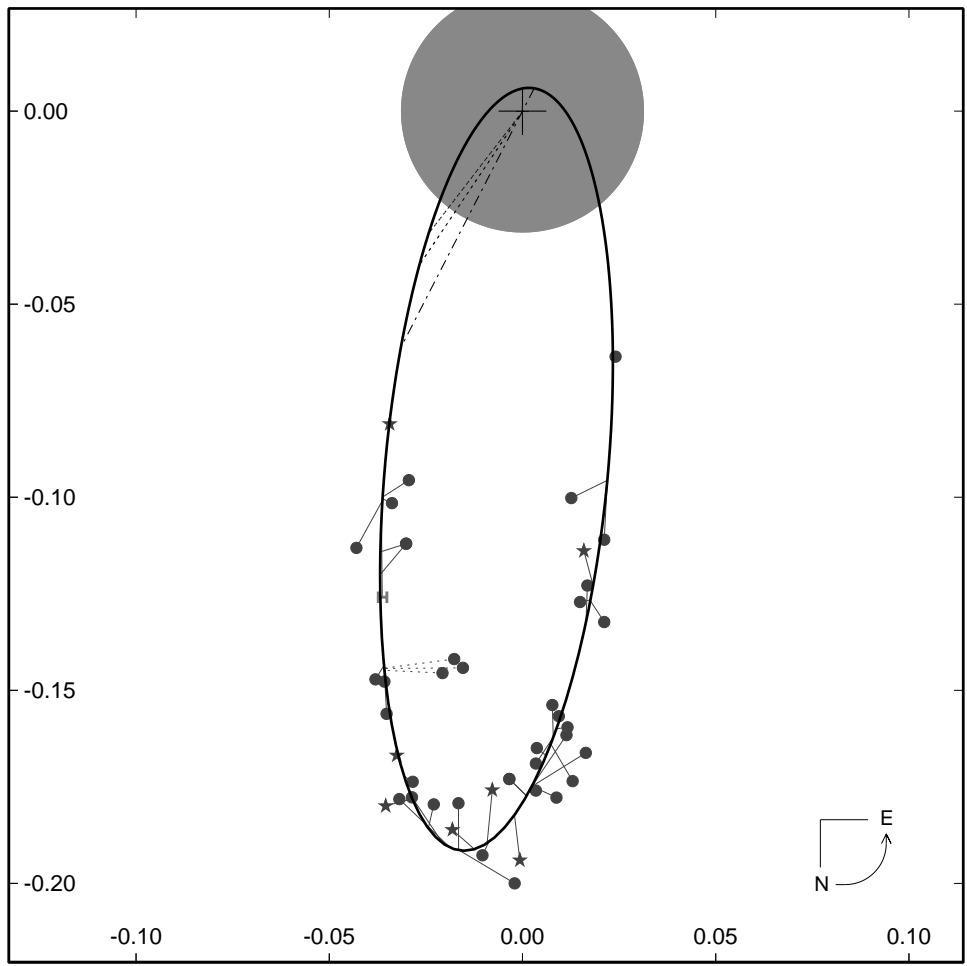}}
\caption[]{New orbit for $\delta$ Sco as described in \S 5.3. The symbols 
have the same meaning as in Fig.~3.}
\label{fig4}
\end{figure}

\placefigure{fig4}
 
\subsection{Notes on Stars Listed in Table 1}

{\bf HD 68243 = WSI  55Ba,Bb:}  This star, $\gamma^1$~Vel, is the B 
component of a group of stars surrounding the bright WR star, $\gamma^2$~Vel
(which is a spectroscopic binary that has been resolved by optical long 
baseline interferometry; North et al.\ 2007).   

{\bf CPD$-$59 2636 = WSI  56:}  A spectroscopic study by Albacete Colombo et
al.\ (2002) detected three spectral components. The brighter star we 
observed probably corresponds to their identification of an A (O7 V) + B 
(O8 V) spectroscopic binary with a period of 3.6~d while the fainter star is
probably their component C (O9 V), itself a single-lined spectroscopic 
binary with a period of 5.05~d. Thus, this is a quadruple system. 

{\bf HD 114737 = WSI  57:} Not detected by Mason et al.\ (1998), it is 
unclear whether the lack of detection earlier was due to the faintness of 
the companion ($\Delta m =1.5$) or to a smaller separation at that time.

{\bf HD 114886 = WSI  58Aa,Ab:} Like HD 114737 above, it is unclear whether 
the lack of detection earlier was due to a magnitude ($\Delta m =1.6$) or 
separation issue.

{\bf HD 124314 = WSI  59Ba,Bb:} This is a close pair associated with the B 
component of the wider known pair COO~167. 

{\bf HD 319703B = WSI  62Ba,Bb:} This is the first measurement of a close 
companion to the B component of the AB pair (separated by $14\farcs5$).
Unfortunately, the A component (also an O-star) was not observed. 

{\bf HD 319718C = WSI  63CD \& CE:} Two additional components were resolved 
while observing the known BC pair. They also can be seen in an {\it HST} 
image made by Ma\'{i}z-Apell\'{a}niz et al.\ (2007) near star B = Pismis 
24-17. Unfortunately, we did not observe the A component = Pismis 24-1 that 
is also a resolved binary (Ma\'{i}z-Apell\'{a}niz et al.\ 2007).

{\bf Cyg OB2-22 = Schulte 22 = WSI  67:}  Our measurements agree with the 
first results on the pair from Walborn et al.\ (2002), who determined O-type
classifications for both components. 

\subsection{Notes on Stars Listed in Table 2}

{\bf HD 47839 = CHR 168Aa,Ab = 15 Mon:} The earlier 15 Mon orbits (Gies et 
al.\ 1993, 1997) are both poor fits to the data listed in Table~2 as well as
other unpublished data from the {\it HST} FGS and optical long baseline 
interferometry. All these data are being collated for a new combined 
solution orbit determination (Gies et al., in preparation).

{\bf HD 97950 = B  1184 AB-F:} This multiple star is actually the core of 
the distant and massive star cluster NGC~3603 (see Fig.~1 in Drissen et al.\
1995). Drissen et al.\ (1995) identify 3 WR stars and 11 O-stars in the core
region. 

{\bf HD 193322 = CHR  96Aa,Ab:} The multiple system HD 193322 was first 
split by speckle interferometry in 1985 (McAlister et al.\ 1987) and 
regularly resolved until closing within the resolution limit of a 4~m 
telescope (30 mas) in 1989. The preliminary 31~y orbit (Hartkopf et al.\ 
1993) had very small residuals but under sampled phase space (covering only
9\% of the orbit). Subsequent to this, the A component was recognized as a 
close 311~d spectroscopic binary (McKibben et al.\ 1998). In addition to the
speckle resolution listed in Table~2, separated fringe packet solutions with
the CHARA Array have been obtained several times since 2005. The ``B'' 
component can act as a calibrator in the field-of-view to allow for rapid 
data acquisition and reduction for a baseline-visibility plus spectroscopy 
combined solution of the inner pair. A preliminary version was recently 
presented (Turner et al.\ 2008a) and a complete analysis of the multiple 
system is underway (ten Brummelaar et al., in preparation) as is 
determination of the distance to the surrounding cluster, Collinder 419
(Roberts et al., in preparation).

{\bf BD$+$40 4212 = ES 1679:} The separation of this binary has declined 
from $\rho= 4\farcs5$ in 1917 (Espin 1918) to $\rho= 3\farcs5$ in 2005.

{\bf WR 146 = NML 1:} Our measurement of this faint pair ($V_{a,b} = 
16.2,16.4$) confirms the discovery observation of Niemela et al.\ (1998).

{\bf WR 147 = NML 2:} The very faint secondary ($V_{a,b} = 15.0,17.2$) is at
the very limit of the USNO speckle camera. This pair was also first resolved
by Niemela et al.\ (1998). Like NML~1 above, this pair was not detected in 
the earlier WR speckle survey of Hartkopf et al.\ (1999) due to the 
limitations of the camera used at that time.

\subsection{Notes on Stars Listed in Table~3}

{\bf HD 103006 = TDS~8073:} The {\it Tycho} satellite (Fabricius et al.\ 
2002) resolved this pair at 0\farcs50 in 1991, but the observation remains 
unconfirmed.

{\bf HD 106508 = FIN 195:} Finsen (1951) resolved this pair at 0\farcs40 in 
1928, and it was measured at 0\farcs34 in 1934 (Rossiter 1955) and 1941 (van
den Bos 1956), and at 0\farcs178 in 1991 (ESA 1997), the only other 
published observation in the last 67 years. Possibly the pair closed to 
$<$0\farcs03 at the time of this observation.

{\bf HD 138923 = FIN 231:} Finsen (1934) resolved this pair in 1929 at a 
separation of 0\farcs18 and followed it over 30 years as it closed to 
0\farcs11 in 1954 and $<$0\farcs119 in 1959 (Finsen 1953, 1954, 1960). No 
published measurements have been made in over 50 years, other than one 
unresolved Hipparcos observation in 1991 (ESA 1997); this suggests the pair 
may have closed to $<$0\farcs03 at the time of this observation.

{\bf HD 152386 = CHR 253:} This object was resolved in 1996 into a 0\farcs55
pair (Mason et al.\ 1998), but the discovery is unconfirmed.

{\bf HD 168878 = CHR 235:} This occultation pair (Africano et al.\ 1978) was
resolved by speckle into a 0\farcs13 pair in 1996 (Mason et al.\ 1996); 
however, this discovery has never been confirmed.

{\bf HD 173524 = ISO 7Aa,Ab:} Isobe et al.\ (1990, 1991) resolved this 
0\farcs20 pair in 1987; however, this discovery has never been confirmed, 
with nine other unresolved observations published to date (Hartkopf et al.\
2001a).

{\bf HD 200595 = BU 1138:} This pair has gradually closed from 0\farcs3 in 
1888 (Hough 1890) to 0\farcs07 in 2001 (Table~2); apparently it closed to 
$<$0\farcs03 at the time of the observation listed in Table~3.

\appendix

\section{Other Systems Observed}

Tables A1 and A2 are identical in form to Tables 2 and 3, respectively, but 
the targets listed here are selected from other sample sets as indicated in 
the more extensive collection of notes. As many of these systems were used 
for either primary (CTIO) or secondary (KPNO) scale and angle calibration, 
many have calculated orbits, with residuals derived from orbital solutions 
given in Table~B1. 

\placetable{taba1}

\placetable{taba2}

\section{Corrected Orbits}

We found that some of those pairs identified as calibration systems in 
Table~A1 (used to investigate differential magnitude detection rates at 
various separations; Fig.~1) had poorly defined orbits. The KPNO measures,
independently calibrated by use of the slit-mask, allowed us to optimize 
these orbits and to generate ephemerides, which helped us calibrate the 
CTIO measures. Going one step further, these CTIO measures could then be 
incorporated in a new, improved orbit solution using the same methodology as
that described in \S5.2. These orbital elements are presented in Table~B1, 
together with their grades (see Hartkopf et al.\ 2001b for a description of 
the grading scale). Also provided in Table~B1 is the reference to the 
previous ``best'' published orbit. Formal errors are listed below each 
element. Future ephemerides are presented in Table~B2 and relative orbit 
plots are illustrated in Figures 5 and 6, with the dashed curve indicating 
the prior orbit listed in Table~B1.

\placetable{tabb1}

\placetable{tabb2}

\begin{figure}[p]
\epsfxsize 4.0in
\centerline{\epsffile{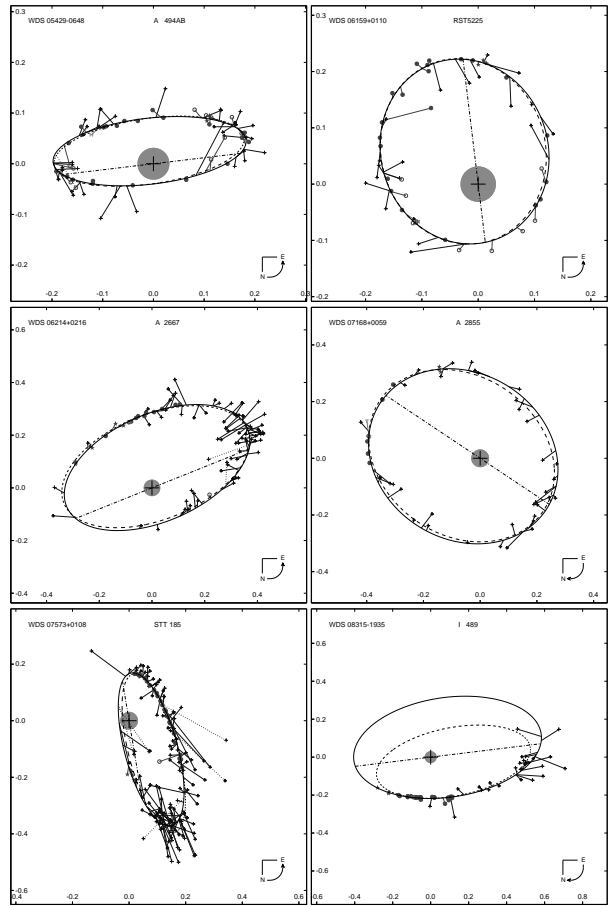}}
\caption[]{New orbits for the systems listed in Table~B1, together with the 
most recent published elements for these systems and all published data in 
the WDS database. See text and Fig.~3 for a description of symbols used in 
this and the following figure.}
\label{fig5}
\end{figure}

\placefigure{fig5}
 
\placefigure{fig6}
 
\begin{figure}[p]
\epsfxsize 4.0in
\centerline{\epsffile{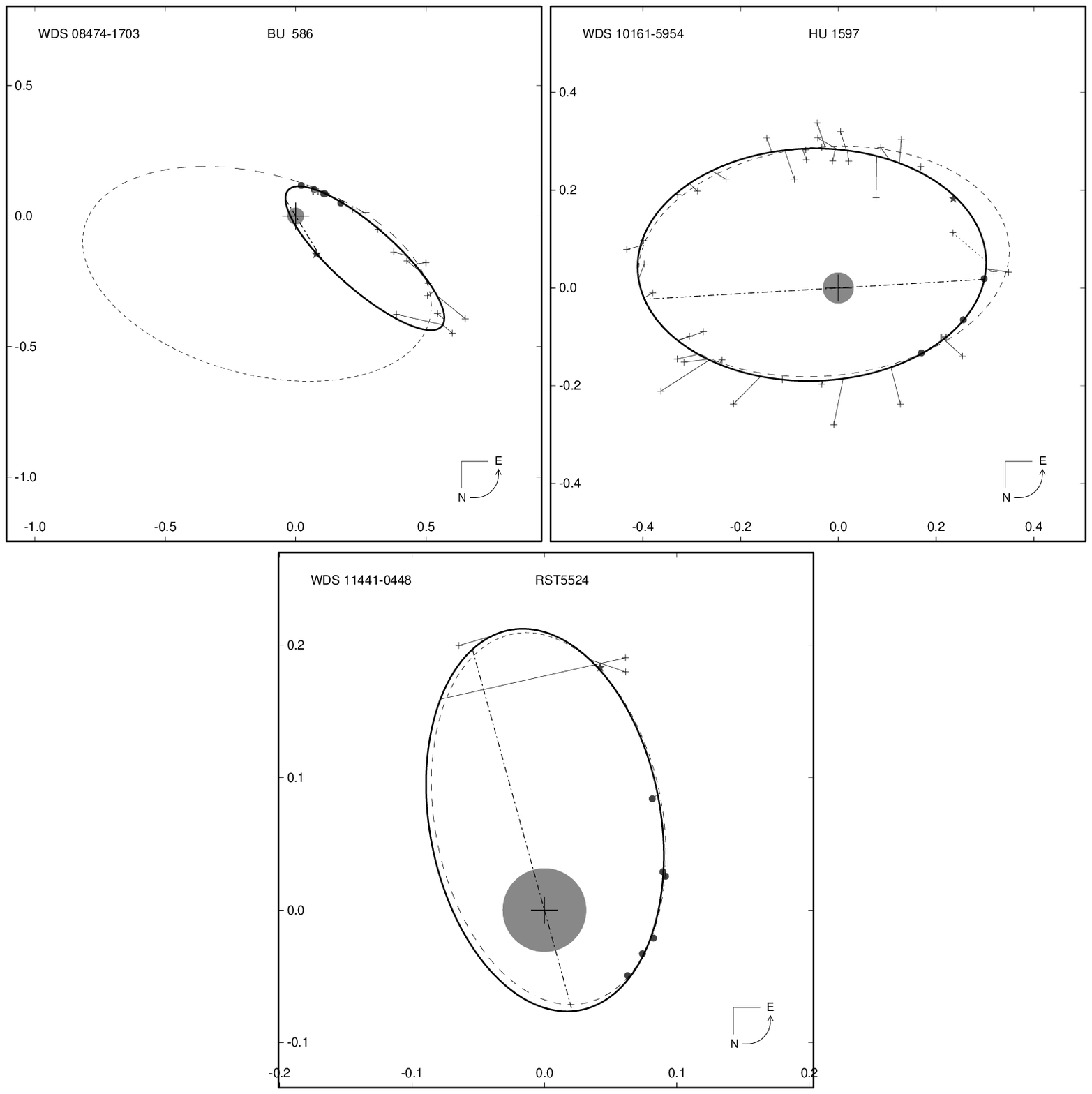}}
\caption[]{New orbits (continued).}
\label{fig6}
\end{figure}


\acknowledgments

The USNO speckle interferometry program has been supported by NASA and the 
Space Interferometry Mission through Key Project MASSIF and is based upon 
work supported by the National Aeronautics and Space Administration under
Grant No. NNH06AD70I issued through the Terestrial Planet Finder Foundation
Science program. This material is based upon work supported by the National
Science Foundation under Grant No.~AST-0506573 and AST-0606861.  This 
research has made use of the SIMBAD database, operated at CDS, Strasbourg, 
France. Thanks are also extended to the U.\ S.\ Naval Observatory for its 
continued support of the Double Star Program. The telescope operators and 
observing support personnel of KPNO and CTIO continue to provide exceptional
support for visiting astronomers. Thanks to Skip Andree, Bill Binkert, Gale
Brehmer, Ed Eastburn, Angel Guerra, Hal Halbedal, David Rojas, Patricio 
Ugarte, Ricard Venegas, George Will, and the rest of the KPNO and CTIO 
staffs. 

\newpage






\begin{deluxetable}{lccccc}
\tabletypesize{\scriptsize}
\tablewidth{0pc}
\tablenum{4}
\tablecaption{Binary Properties of Galactic O-Stars\label{tab4}}
\tablehead{
\colhead{Star} &
\colhead{Spectro.} &
\colhead{No.\ Vis.} & 
\colhead{} & 
\colhead{}   &
\colhead{Spectroscopic} \\
\colhead{Name} &
\colhead{.Status} &
\colhead{Comp.} & 
\colhead{Environment} & 
\colhead{Runaway?}   &
\colhead{Reference} 
}
\startdata
HD 164019        &C    & 1 &Field:                  &no:&1957MmRAS..68....1F\\
HD 162978        &C    & 1 &Sgr OB1                 &no &1980ApJ...242.1063G\\
HD 168941        &C    & 1 &Field:                  &yes&1957MmRAS..68....1F\\
Herschel 36      &U    & 1 &NGC 6530,Sgr OB1        &no &1997AJ....113..823R\\
9 Sgr            &SB2? & 1 &Sgr OB1                 &no &2002A\&A...394..993R\\
HD 164816        &SB2? & 1 &Sgr OB1                 &no &1996ApJ...463..737P\\
HD 165052        &SB2O & 1 &Sgr OB1                 &no &2007A\&A...474..193L\\
$\zeta$ Oph      &C    & 1 &Sco OB2                 &yes&2005ApJ...623L.145W\\
\enddata
\end{deluxetable}



\begin{deluxetable}{lr}
\tablenum{6}
\tablewidth{0pt}
\tablecaption{Orbital Elements for $\delta$ Ori = HEI~42\label{tab6}}
\tablehead{
\colhead{Element} & 
\colhead{Value} 
}
\startdata
$P$ (y)        &  201 \nl
$a$ ($''$)     & 0.26 \nl
$i$ (deg)      &  108 \nl
$\Omega$ (deg) &  139 \nl
$T$ (BY)       & 1957 \nl
$e$            & 0.56 \nl
$\omega$ (deg) &  236 \nl
\enddata
\end{deluxetable}


\begin{deluxetable}{lcccc}
\scriptsize
\tablenum{7}
\tablewidth{0pt}
\tablecaption{Orbital Elements for $\delta$ Sco = LAB~3\label{tab7}}
\tablehead{
\colhead{} & 
\colhead{Bedding} & 
\colhead{Hartkopf et al.} & 
\colhead{Miroshnichenko et al.} & 
\colhead{} \nl
\colhead{Element} & 
\colhead{(1993)} & 
\colhead{(1996)} & 
\colhead{(2001)} & 
\colhead{This work} 
}
\startdata
$P$ (y)        & 10.5   & $10.583 \pm 0.075$  & 10.58\tablenotemark{a} & $10.68 \pm 0.05$          \nl
$a$ ($''$)     & 0.11   & $0.1067 \pm 0.0067$ & 0.107\tablenotemark{a} & $0.104 \pm 0.006$         \nl
$i$ (deg)      & 70     & $48.5 \pm 6.6$      & $38 \pm 5$             & $39 \pm 8$                \nl
$\Omega$ (deg) & 0      & $159.3 \pm 7.6$     & 175                    & $153 \pm 9$               \nl
$T$ (BY)       & 1979.3 & $1979.41 \pm 0.14$  & $2000.693 \pm 0.008$   & 2000.693\tablenotemark{b} \nl
$e$            & 0.82   & $0.92 \pm 0.02$     & $0.94 \pm 0.01$        & 0.94\tablenotemark{b}     \nl
$\omega$ (deg) & 170    & $24 \pm 13$         & $-1 \pm 5$             & $29 \pm 12$               \nl
\enddata
\tablenotetext{a}{Parameter adopted from Hartkopf et al.\ (1996) solution.}
\tablenotetext{b}{Parameter adopted from Miroshnichenko et al.\ (2001) solution.}
\end{deluxetable}%

\end{document}